\documentclass{aa}
\usepackage{psfig}
\title{Lensing galaxies: light or dark?}
\author{N.~Jackson\inst{1} \and P.~Helbig\inst{1} \and I.~W.~A.~Browne\inst{1}
\and 
C.~D.~Fassnacht\inst{2} \and L.~Koopmans\inst{3} \and  D.~Marlow\inst{1} \and 
P.~N.~Wilkinson\inst{1}}
\offprints
{N.~Jackson
}
\mail{njj@jb.man.ac.uk}
\institute{University of Manchester, NRAL, Jodrell Bank, 
Macclesfield SK11 9DL, UK \and 
California Institute of Technology, Mail Code 105-24,
Pasadena,  CA 91125, USA \and
Kapteyn Astronomical Institute, Postbus 800, 9700 AA
Groningen, The Netherlands}
\date{
accepted
}

\thesaurus{01     % A&A Section 6: Form. struct. and evolut. of stars
              (11.06.2; 11.19.7; 12.04.1; 12.07.1 ) }

\authorrunning{N.~Jackson et al.}
\begin{document}
\maketitle
\begin{abstract}

In a recent paper, Hawkins (1997) argues on the basis of statistical
studies of double-image gravitational lenses and lens candidates that
a large population of dark lenses exists and that these outnumber
galaxies with more normal mass-to-light ratios by a factor of 3:1. If
correct, this is a very important result for many areas of astronomy
including galaxy formation and cosmology. In this paper we discuss our
new radio-selected gravitational lens sample, JVAS/CLASS, in order to
test and constrain this proposition. We have obtained ground-based and
HST images of all multiple-image lens systems in our sample and in 12
cases out of 12 we find the lensing galaxies in the optical and/or
near infrared.  Our success in finding lensing galaxies creates
problems for the dark lens hypothesis. If it is to survive, ad hoc
modifications seem to be necessary:  only very massive galaxies
($M\ga9\times10^{11}{\rm M}_{\odot}$) can be dark, and the cutoff in
mass must be sharp. 
Our finding of lens galaxies in all the JVAS/CLASS systems is
complementary
evidence which supports the conclusion of Kochanek et al. (1997) that 
many of the wide-separation optically-selected pairs are physically 
distinct quasars rather than gravitational lens systems.

\keywords{Galaxies: fundamental parameters -- cosmology: dark matter --
gravitational lensing}

\end{abstract}

\section{Introduction}

Gravitational lensing is an important phenomenon because it probes mass
distributions of galaxies independently of their optical luminosity. In
particular, it provides a unique way to search for galaxies with very high
mass-to-light ratios which are otherwise difficult to detect.

In a recent paper, Hawkins (1997) performs such a search, based on
published data on eight two-image gravitational lens systems and
candidate lens systems with image separations
$>2^{\prime\prime}$. Statistical arguments are presented that these
are indeed genuine lens systems rather than chance associations of
unrelated quasars. For example, several of the quasar pairs have more
similar colours than one would expect to find in the general
population of quasars, and in at least one case that the spectra are
so similar that when one spectrum is divided by the other, the result
is a constant ratio to within the noise (Hawkins et
al. 1997). No lensing galaxies are found in six of the
systems, resulting in mass-to-light ratios of up to 22000 in the most
extreme case. The inference is that the lensing is done by
underluminous `dark galaxies' with very substantial components of
dark matter, with serious implications for cosmology as well as lensing
studies. A further paper by Jimenez et al. (1997) discusses how
such galaxies could be formed. However, 
Kochanek, Falco \& Mu\~{n}oz (1997) have argued against a lensing
interpretation and for the hypothesis that the Hawkins lenses are binary
quasar pairs, based on existing statistics of large separation lenses,
because a population of wide-separation, optically-selected lenses
should imply a significant, unseen, population of corresponding
radio-selected lenses. They also discuss formation scenarios for such
binary quasar pairs. 
In this paper we describe a new, well-defined, sample of gravitational
lens systems selected from the JVAS/CLASS radio surveys.  We discuss
the search for the lensing galaxies in the 12 systems found so far and
assess the implications for the proposal that dark lensing galaxies
exist. In Section 2 we give a brief description of JVAS/CLASS and list
the confirmed lens systems. In Section 3 we summarize the
observations, both with ground-based telescopes and with the HST, in
which we detect lensing galaxies in all 12 systems. Finally, in
Section 4, we present our conclusions and discuss further the question of
whether the optically-selected pairs (Hawkins 1997) are lens systems,
physically unrelated quasar pairs or related quasar pairs.

\section{The JVAS/CLASS radio lens searches}

The objective of the JVAS/CLASS surveys is to observe all
northern-hemisphere sources with flux densities $>$30~mJy at 5~GHz,
and with a radio spectral indices flatter than $-$0.5
($S_{\nu}\propto\nu^{\alpha}$). The brighter section of the survey,
the Jodrell Bank-VLA Astrometric Survey (JVAS) consists of the
brightest $\sim$2400 sources ($S_{\rm 5GHz}>200$~mJy) and is described
by Patnaik et al. (1992a) and Wilkinson et al.
(1998) . The Cosmic Lens All Sky Survey (CLASS), which is a
collaboration between groups at Jodrell Bank, Caltech, Dwingeloo and
Leiden, is an extension of this work to weaker sources. The surveys
used to define the JVAS/CLASS sample, and determine the radio spectral
indices, are the NRAO 5-GHz surveys at the high-frequency end (Gregory
\& Condon 1991; Gregory et al. 1996) and either the Westerbork
Northern Sky Survey (WENSS) at 325$\,$MHz (de Bruyn et al., in preparation) or
the Texas 365$\,$MHz catalogue (Douglas et al. 1996) at the low
frequency end. So far, approximately 9000 sources have already been
observed of which about 8000 have been detected with 8.4~GHz flux
densities $\geq$ 20~mJy.

Every effort is being made to ensure that the finished survey is
complete in the sense that all gravitational lenses with image
separations $<6^{\prime\prime}$, and with image flux ratios $<$10 will be
detected. This task is made easier by the fact that flat-spectrum
radio sources have intrinsically simple structures (mostly
unresolved on the arcsecond scale). Initial observations are made in
snapshot mode with the VLA in A-configuration at 8.4GHz.  These give
0.2~arcsec-resolution images with dynamic range $\geq$100 on the
brightest component. Any object which consists of more than one
component with angular separation 0\farcs3$<s<6$~arcsec
between the components is followed up by higher-resolution radio
observations using some combination of the VLA, MERLIN and the
VLBA. In this phase of the search for lens systems, candidates are
rejected if the images have very different radio spectra, if the
images have very different percentage polarizations, or if the high
resolution maps reveal extended structure inconsistent with lensing
(most often the putative lensed images are shown to have very
different surface brightnesses). More details of the lens search
procedure are given by King et al (1998a).

The radio follow-up observations of the candidates from all the
sources so far observed with the VLA are virtually complete. So far we
have twelve confirmed lens systems which we list in Table 1. {\em 
There are no systems that have passed all the radio
tests which have been subsequently rejected by optical 
observations}. Hence we take as a confirmed system as one which has
passed the radio tests; here we discuss the subset of these which have
so far been observed with HST. 
There remain five recently-discovered good candidates
(not included in Table 1) which still require
optical follow-up observations.
These five candidates have separations $<3$~arcsec.

\begin{table*}
\caption[]{Radio-loud lensed systems from JVAS/CLASS, in order of
decreasing image separation. In column 2 we give the reference to the
discovery paper: column 3 gives the number of images
of each source component.  z$_{\rm s}$ (column 4) and z$_{\rm l}$ (column 5)
are the source and lens redshift. Redshifts for
three systems are taken from Fassnacht \& Cohen (1998) and the
lens galaxy redshift for B1422+231 is from Kundi\'c et al. (1997).  The
image separation (column 6) is given in arcsec.  In columns 7-9, the
lens magnitudes (Johnson-Cousins V, I, CIT H; typical errors $\pm$0.2
magnitudes from measurements of different frames) and mass-to-light ratios 
are given. Magnitudes and inferred lens masses given are within the Einstein
radius, and 
are from our own data, except for B1422+231 (Impey et al. 1996),
B0218+357 (fits by Hjorth 1997: the magnitudes are
total magnitudes rather than within the Einstein radius, making
the inferred $M/L$ ratio slightly too small), B1030+074 
 (Xanthopoulos et al. 1998).  $M/L$ ratios for 6 objects, plus other
non-JVAS/CLASS objetcs, are given by Keeton, Kochanek \& Falco (1997). 
The mass-to-light ratios ({\bf bold
type}) have no K- or evolution corrections (see the text for a
discussion of these) and are in units of ${\rm M}_{\odot}/{\rm
L}_{\odot}$. In column 10 we give the lens mass in units of
10$^{10}$M$_{\odot}$. MG0414+054 is a rediscovery of a lens system
from the MIT-Greenbank Survey.  B2114+022 has 4 compact
components, of which at least two are gravitational images; it is not
clear if all four components are images of the same object;
two galaxies are associated with the lens (Augusto et
al. 1998, in preparation), as is also the case in B1127+385. 
References: 1. Augusto et al. 1998; 2. Hewitt et
al. 1992; 3. Myers et al. 1995; 4. Fassnacht et al. 1998;
5. Xanthopoulos et al. 1998; 6. Jackson et al. 1995; 7. Patnaik et
al. 1993; 8. Jackson et al. 1998; 9. Sykes et al. 1998; 10. King et
al. 1997, King et al. 1998; 11. Koopmans et al. 1998; 
12. Patnaik et al. 1992.}

\begin{tabular}{llllllllll}
\hline 
Source  & Ref & No. of & Source   &  Lens & Max.  & 
\multicolumn{3}{c}{Magnitude/{\bf M-to-L ratio}}&Mass\\
        &    & images & redshift z$_{\rm s}$  & redshift z$_{\rm l}$
&separ$^{\rm n.}$&V & I & H &/10$^{10}$M$_{\odot}$\\
\hline
B2114+022  &1& 2,4?&  -     & 0.32,0.59   & 2.57  &    & 
&17.0,17.2/{\bf 2.9,1.3}&26,52\\
MG0414+054 &2 &4&  2.62  &  -    & 2.09   &    &   
&18.2/{\bf 2.5}&24\\
B1608+656  &3&  4&  1.39  & 0.64  & 2.08   &21.4/{\bf
17.9}&19.0/{\bf 6.4}& &52\\
B2045+265  &4&  4& 1.28  &  0.87   & 1.86    &    &  &19.0/{\bf 
6}&86\\
B1030+074  & 5&  2& 1.53   & 0.599 & 1.56   &22? &20.4/{\bf 7.3} 
&&22\\
B1600+434  &6&  2&  1.57  & 0.415  & 1.39   &23? &21.2/{\bf
17.6}&18.5/{\bf 2.4}&12\\
B1422+231  & 7&  4&  3.62  & 0.337  & 1.28   &21.4/{\bf 16}&&&12\\
B0712+472  &8&  4&  1.34  & 0.406 & 1.27   &22.4/{\bf
20.5}&20.0/{\bf 4.7}&17.7/{\bf 0.9}&9\\
B1933+507  &9& 10&  -     & 0.755 & 1.17   &$<$23.5/{\bf
17.2}&21.8/{\bf 7.4}&18.6/{\bf 0.6}&10\\
B1938+666  &10&  4&  -     &  -    & 0.93   &    &  &18.5/{\bf 
0.8}&6\\
B1127+385 & 11 & 2 & - & - & 0.70 & 24.4,25.5/
& 22.5,23.5/& & 2.7,1.1\\
          &    &   &   &   &      &{\bf 17.3,18.6}&{\bf 6.2,6.1}&&\\
B0218+357  &12&   2&   0.96  & 0.685  & 0.334 &22.1/{\bf
1.3}&20.0/{\bf 0.4}& &2\\
\hline
\end{tabular}
\end{table*}

In Figure 1 we show the histogram of the maximum image separations
given in Table 1 for the twelve confirmed lens systems. Our search
technique is designed to pick up systems with image separations in the
range 0.3~arcsec to 6~arcsec, and we believe it to be complete
in this range. In particular, we have found no lens systems with
separations $>3$~arcsec amongst 8000 radio sources mapped.
However, three systems have separations between 2~arcsec and
3~arcsec and hence lie in the separation range discussed by
Hawkins.
%of which two are single-galaxy lenses.
%In Table 1 we show the redshifts and image separations of the
%lenses, together with their optical and infrared magnitudes and
%inferred mass-to-light ratios.

\begin{figure}
\psfig{figure=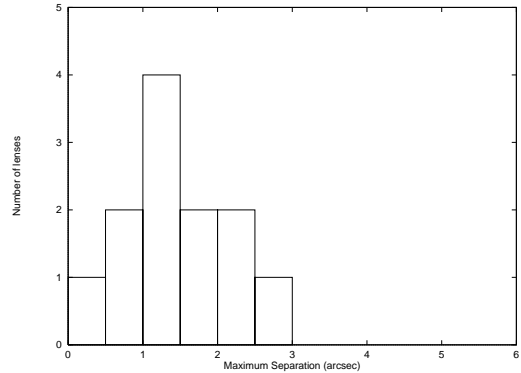,width=7cm,angle=-90}
\caption[]{Histogram of the maximum image separation (in arcsec) for the 12 confirmed JVAS/CLASS lens systems.}
\end{figure}

\section{Optical and infrared follow-up; lensing galaxies}

All the systems listed in Table 1 have been observed with the HST,
either with WFPC2 (in V and I-bands) or with NICMOS (in H-band). Full
details of the HST observations will be presented elsewhere. 
Figure 2 shows an example of one of
the lensing galaxies detected, that of B1608+656.  In all but one case
the lensing galaxy has been detected in a position consistent with the
gravitational lens hypothesis, the exception being B2114+022 where two
galaxies are detected (Augusto et al. 1998). The resulting magnitudes
are listed in Table 1. We give the magnitude within the Einstein
radius except for B0218+357 and B1422+231 where values for total
luminosity  are taken from
the literature.  Multiple optical counterparts corresponding to two or
more of the radio images are detected in all systems except
B2114+024.

\begin{figure}
\begin{tabular}{c}
\psfig{figure=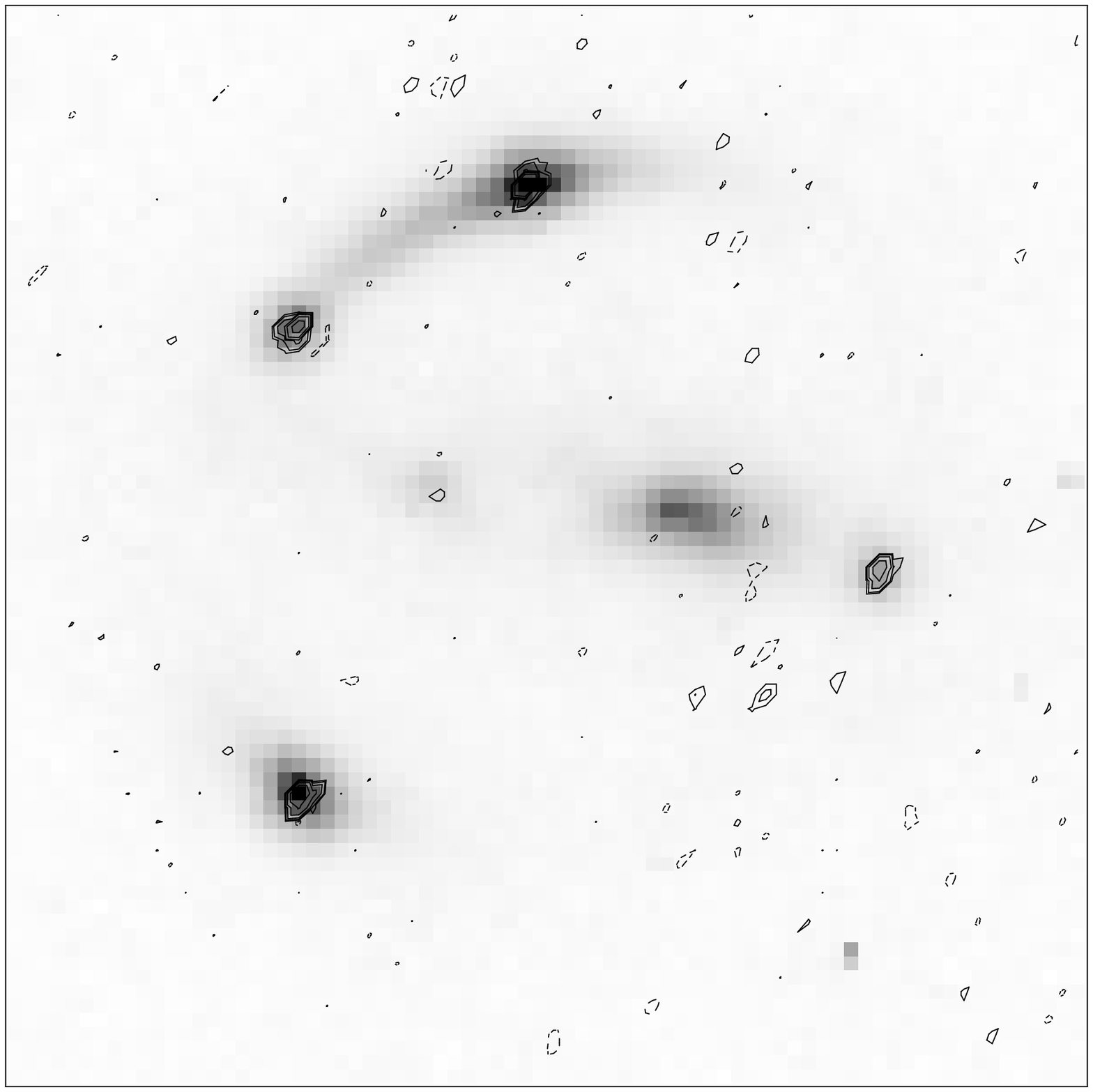,width=5.1cm}
\end{tabular}
\caption{The MERLIN 5-GHz image (contours) of B1608+656 
(made from a 12-hour track on 1996 January 29; noise level
of 130$\mu$Jy/beam; contours are at
300$\mu$Jy/beam$\times$[$-$1,1,2,4,8,16,32]) superimposed on 
the HST/WFPC2/814-nm image (greyscale) 
of B1608+656. The lensing galaxy is clearly visible in
the optical image, as are lensed arcs due to the extension of the
lensed image, which is a post-starburst galaxy (Fassnacht et
al. 1996).}
\end{figure}

In most cases the redshifts of the lensing galaxy and the lensed
object are determined and this enables us to estimate a lens mass. In
the cases where the redshift information is incomplete we assume a
lens redshift of 0.5 and a source redshift of 1.5. In each case we
have adopted a singular isothermal sphere model for the mass
distribution of the lens. The mass-to-light ratios so
calculated\footnote{We assume $H_{0}$=60\,km\,s$^{-1}$\,Mpc$^{-1}$,
$\Omega_{0}$=1, $\lambda_{0}$=0} are given in Table 1, and are all
consistent with the values expected if the lenses are normal luminous
elliptical galaxies ($M/L_{R}\leq\sim$20, e.g. Oegerle \& Hoessel 1991).
Both mass and light are estimated within the Einstein radius: the
(non-SIS) correct mass distribution to adopt outside this radius is
beyond the scope of this paper.

No K-corrections or evolutionary corrections have been applied to give
the values presented in Table 1. However, these corrections are
relatively unimportant in the present context as for an E galaxy at
redshift 0.6 the combined value of these corrections give correction
factors of $\times$0.35, $\times$1.15 and $\times$1.85 in V, I, and H,
respectively, for the mass-to-light ratios (Poggianti 1997; the
correction factors for spirals are not very different). It can be
seen, therefore, that the corrected mass-to-light ratios for our
lensing galaxies range from $\sim$1 to around 20. Of the two systems
with the largest $M/L$, B1600+434 is an edge-on spiral (Jaunsen \&
Hjorth 1997; Koopmans, de Bruyn \& Jackson 1998) and some internal
reddening is therefore expected, which will boost the inferred mass-to-light
ratio. The other high-$M/L$ object is B2045+265, which
has a large separation and contains a relatively faint (but not dark)
lensing galaxy.

\section{Discussion}

%\subsection{The global population of lenses}

In our radio-selected lenses, even without the detection of
the lensing galaxy, there can be no doubt that one is dealing with
gravitational lens systems, because of radio spectral index and
radio/optical similarities of the images.  The results we have obtained 
suggest strongly that lensing galaxies have $M/L_{I}\la50$ and thus are not
dark.  Twelve out of twelve of the JVAS/CLASS lensed systems have
$M/L_{I}$ ratios suggestive of normal luminous galaxies. Even if {\em all}
the remaining 5 candidates were lensed systems, and {\em all} of these
had no detectable lensing galaxy, this would still leave over 60\% of
lens systems with detected, and relatively normal, lensing galaxies
compared to $<$25\% reported by Hawkins (1997).

There is a possible explanation for the difference between our results
and those of Hawkins. This is that the Hawkins sample has a
fundamentally different distribution of image separations, as it contains only
systems with separations greater than 2$^{\prime\prime}$, which, since
the separation $s$ is a function of lens mass $M$ 
(e.g. Schneider et al. 1992), corresponds to relatively massive lensing
galaxies  ($\ga9\times10^{11}{\rm M}_{\odot}$, see Table 1). 
Could only massive galaxies be dominated by dark matter?

%\subsection{Large-separation systems}

Let us concentrate, therefore, on radio lensed systems with separation
$>$2~arcsec. We will further restrict our consideration to those
incontrovertible lens systems in which optical counterparts to the
radio components have been detected.  There are two such systems with
separation $>$2~arcsec in the JVAS/CLASS sample\footnote{We do not
include B2114+024 as there is no detection of optical counterparts to
any of the radio images.}.
There are two further known radio-selected systems with
$s>2^{\prime\prime}$; B0957+561 (from Hawkins' sample) and B2016+12
(Lawrence et al. 1984). In all four of these normal (not dark) lensing
galaxies are detected. Thus amongst {\em confirmed} lens systems with
separations $>2^{\prime\prime}$ there is no evidence for dark lenses.

%\subsection{Four-image systems}

There is no reason why any bias should be introduced by the use of
radio selected systems; it is simply a way of securely identifying lens
systems without the necessity of identifying a
lensing galaxy. The discrepancy between the results for radio-selected
lens systems, even for large separation lenses, and the mostly
optically-selected sample of Hawkins (1997) is therefore puzzling.  

We note, however, that Hawkins only considers systems with 2 images.
Among the radio-selected systems there are many 4-image
systems. Could the dark lenses be found only in the 2-image
systems?  Kochanek et al. (1997) pointed out that in general, one 
would expect a greater preponderance of
4-image systems where the lensing mass distribution, particularly the
halo, is more elliptical (e.g. King \& Browne 1996), and that one might
therefore postulate that `dark lenses' have spherically symmetric mass
distributions.  This, however, fails to account (Kochanek et al. 1997) 
for why the existing radio-selected two-image systems do not have dark 
lenses, as well as for why we find no two-image systems with dark lenses
in JVAS/CLASS.

%\subsection{Summary: implications for dark lenses}

In summary, there are two conditions that must be met in order to reconcile our
observations with the idea of dark lenses. First, in order for us to
find no dark lenses in the JVAS/CLASS sample, there must be a sharp
cutoff such that all dark lenses must have masses of
$\ga9\times10^{11}{\rm M}_{\odot}$. Second, in order to explain the
detection of lenses in the radio lens systems with separations
$>$2~arcsec (mostly quads), dark lenses should have much rounder
mass distributions than the luminous lenses (Kochanek et al. 1997). 
The alternative to these
two conditions is to suppose that most, or even all, of the systems
discussed by Hawkins in which no lensing galaxy is found, are indeed
not lens systems. This is despite the fact that in each individual
case we find the arguments persuasive that the quasar pairs are not
random associations.  However, if there are no dark lenses, the
frequency of close physically associated pairs of quasars has hitherto
been greatly underestimated.

\acknowledgements{We acknowledge the help of our collaborators on the
JVAS and CLASS lens search. This research was supported in part by the 
European Commission, TMR Programme, Research Network Contract ERBFMRXCT96-0034
`CERES', and was based on observations with the Hubble Space Telescope,
obtained at the Space Telescope
Science Institute, which is operated by Associated Universities for
Research in Astronomy Inc. under NASA contract NAS5-26555. MERLIN is
operated as a National Facility by the University of Manchester, NRAL,
on behalf of PPARC.}

\end{document}